\documentclass[final, 3p, times]{elsarticle}

\usepackage{graphicx}
\usepackage{amssymb}
\usepackage{amsmath}
\usepackage{booktabs}
\usepackage{multirow}
\usepackage{multicol}
\usepackage{ulem}

\begin{document}

\begin{frontmatter}

\title{Beat Frequency Induced Transitions in Synchronization Dynamics}

\author[1]{Gabriel Marghoti}
\author[1,2,3]{Thiago L. Prado\corref{cor1}}\ead{thiago@fisica.ufpr.br}
\author[3]{Miguel A.F. Sanju\'{a}n}
\author[1,2]{Sergio R. Lopes}

\address[1]{Departamento de F\'{i}sica, Universidade Federal do Paran\'{a}, Curitiba, Paran\'{a}, Brazil}
\address[2]{Interdisciplinary Center for Science, Technology and Innovation CICTI, Universidade Federal do Paran\'a, Curitiba, Brazil}
\address[3]{Nonlinear Dynamics, Chaos and Complex Systems Group, Departamento de
F\'{i}sica, Universidad Rey Juan Carlos, Tulip\'{a}n s/n, 28933 M\'{o}stoles, Madrid, Spain}

\cortext[cor1]{Corresponding author.}

\begin{abstract}

In neurosciences, the brain processes information via the firing patterns of connected neurons operating across a spectrum of frequencies. To better understand the effects of these frequencies in the neuron dynamics, we have simulated a neuronal network of Izhikevich neurons to examine the interaction between frequency allocation and intermittent phase synchronization dynamics. As the synchronized population of neurons passes through a bifurcation, an additional frequency mode emerges, enabling a match in the mean frequency while retaining distinct most probable frequencies among neurons. Subsequently, the network intermittently transits between two patterns, one partially synchronized and the other unsynchronized. Through our analysis, we demonstrate that the frequency changes on the network lead to characteristic transition times between synchronization states. Moreover, these transitions adhere to beat frequency statistics when the neurons' frequencies differ by multiples of a frequency gap. Finally, our results can improve the performance in predicting transitions on problems where the beat frequency strongly influences the dynamics.

\end{abstract}


\end{frontmatter}

\section{\label{sec:intro}Introduction}

Synchronization phenomena have been studied in various scientific fields such as physics, chemistry, and biology, as a universal concept of dynamical systems \cite{Pikovsky_SYnchronizatioon_a_universal_concept, Boccaletti_Sync_book}. Moreover, in neurosciences, the synchronization of neuronal activity is crucial for the proper functioning of the brain. Neurons communicate with each other by generating rhythmic activity patterns, and the synchronization of these patterns plays a crucial role in information processing and cognitive functions \cite{language_brain_rhythms_30Hz, jones2016brain_rhythms_20hz}. However, when neurons have different natural frequencies, phase synchronization can be challenging to achieve \cite{buzsaki2006rhythms,guevara2017neural_sinc}. 

In the hippocampus, a brain region considered essential for memory, clusters of neurons exhibit distinct frequencies. 
These cluster differences are essential to select different attributes from the synaptic current and consequently to information processed locally by distinct neurons \cite{pike_distinct_freq_2004,hyafil2015neural}. Moreover, certain neurological disorders like Parkinson's disease and epilepsy are associated with abnormal synchronization patterns in neuronal networks \cite{hammond2007pathological, wang2020brief, rubchinsky2012intermittent}. Understanding phase synchronization could contribute to a better knowledge of these pathological dynamic patterns and potentially inform therapeutic strategies for such disorders.

Mathematical modeling of neurons is typically done by using differential equations. The Hodgkin-Huxley model is one of the most known, but others have been proposed over time. Some models are directly based on the Hodgkin-Huxley model and implement additions (e.g., Huber-Braun model \cite{braun1999low_rato_burst}) or simplify the base model (e.g., Fitz\-Hugh-Nag\-u\-mo model \cite{Fitzhugh_nerve_membrane, Nagumo_circuit}). There are other models which aim to mimic the dynamical behavior but with almost no physiological connection (e.g., Mo\-rris\--Le\-car \cite{Morris_Lecar} and Hindmarsh\-–Rose \cite{Hindmarsh_Rose}). 

Furthermore, it is possible to use difference equations, also called map-based neuron models \cite{sanjuan_neuronMap2011, CHIALVO1995461, Rulkov_Map} that have been built inspired by continuous models. Another class of models presents a mixture of continuous and discrete rules for temporal evolution using discontinuous differential equations.
Such systems are used to calculate neuron firing patterns, acquiring a wide range of realistic dynamic behaviors with low computational costs \cite{izhikevich2008_classes_neurons, Firing_patterns_integrate_and_fire_Naud}.

The Izhikevich neuron model \cite{Izhikevich_model_2003} presents a balance between computational efficiency and the capacity to replicate diverse biological firing patterns exhibited by neurons. Extensive bifurcation analysis for one Izhikevich neuron has been performed in Refs.~\cite{izhikevich2007_book, tamura2009bifurcation_Izhik}, while some other studies explore the dynamics of heterogeneous neurons around the period-adding bifurcation \cite{izhikevich2007_book, fox2022bursting} and also the mean field derivation \cite{Jieqiong_mean_field_derivation}.

We show that the activity of individual neurons is crucial in determining the duration of synchronization patterns in a network due to the frequency distribution generated within it. This phenomenon can be associated with the beat period of a group of oscillators, such as the \textit{pendulum waves} \cite{berg1991_pendulum_wave, flaten2001pendulum}. We also link the transition times to the \textit{missing fundamental illusion}, a psycho-acoustic effect where the brain perceives a pitch frequency with no source vibration. This effect has been studied {in a more general framework for oscillating systems called \textit{Ghost Stochastic Ressonance} phenomenon \cite{chialvo2003we, balenzuela2012ghost, chialvo2002subharmonic}.

Our study aims to investigate the duration of the intermittent partial phase synchronization dynamics within clusters of oscillators described by \citet{marghoti2022intermittent}. More specifically, how characteristic times in coupled neuronal oscillators affect intermittent synchronization transitions.

This work introduces the Izhikevich neuronal network model in Sec.~\ref{sec:model}. Next, it explores the isolated neuron and the overall network dynamics in Sec. ~\ref{sec:isolated_dynamics} and Sec. \ref{sec:net_dyn}, respectively. Finally, Sec. \ref{sec:beat_formulation} and Sec. \ref{sec:transitions} explain the intermittent dynamics of the network based on individual neuron properties and the beat frequency formalism.

\section{\label{sec:model}The Neuronal Model}

The Izhikevich model is defined by a set of differential equations and a reset rule. The differential equations represent the time evolution of the neuron membrane potential $v(t)$ and recovery variable $u(t)$ \cite{Izhikevich_model_2003, izhikevich2007_book}.
{This neuronal model efficiently simulates various firing patterns seen in biological neurons. Here, we focus on a pattern marked by a sequence of spikes followed by a quiescent period, known as bursting dynamics, observed in the hippocampus \cite{cooper2005output}, basal ganglia \cite{cagnan2019temporal}, and dopaminergic neurons in the midbrain \cite{gonon1988nonlinear}. However, it's important to note that neuronal heterogeneity within cell types is expected in the brain \cite{Cembrowski2019}.}

The neuronal network model we use encompasses heterogeneous bursting dynamics, through the distribution of a control parameter $a$, specifically the effects of spike gain in each burst due to a bifurcation in the local dynamics. We simulate a heterogeneous network of Izhikevich neurons that differs in their parameters $a$, so that the equations of motion for the $i$th neuron in the network are given by\\
\begin{equation}
\dot{v_i}(t) = 0.04v_{i}^{2}(t)+5.00v_i(t)+140-u_i(t)+I(t),\\    
\label{eq:v_izhik}
\end{equation}
\begin{equation}
\dot{u_i}(t) = a_i[bv_i(t) - u_i(t)].
\label{eq:v_izhik2}
\end{equation}
The reset rule can be written as follows:
\begin{equation}
\begin{split}
& \mathrm{if\,} v_i(t) \geq 30.0 \ \mathrm{mV}\mathrm{,} \\ & \mathrm{then\,} \left[ v_i(t) \leftarrow c \right] \mathrm{\,and\,} \left[ u_i(t) \leftarrow u_i(t) + d \right], \label{reset}
\end{split}
\end{equation}
where $i=1, 2, ...,N$, and over dots represent time derivatives.

The term $I(t)$ is the stimulus each neuron receives, which is in general a sum of several processes \cite{bennett2004electrical_gap_junction,pereda2014electrical_chemical}.  However, to keep the model simple and to isolate the possible cause of complex dynamics to the parameter distribution for the neurons,} we assume a synaptic current proportional to the network mean-field \cite{Izhikevich_model_2003, izhikevich2007_book}. Therefore, we simulate a densely connected small group of distinct neurons, so that the synaptic input for all neurons is
\begin{equation}
    I(t) = I_b + \gamma \langle v \rangle (t),
    \label{eq:I}
\end{equation}
where $I_b$ is the input bias related to the constant extra-cellular electric potential bias and $\gamma$ is the coupling strength that mediates  $\langle v \rangle (t)$ over the individual neuron. We notice the input $I(t)$ for each neuron is time-dependent due to the mean effect of the other neurons, therefore the dynamics of the neural network is governed by the interactions and feedback among its constituent neurons rather than by external inputs.

Parameters $a_i$ are distributed between $a_{min} = 0.013$ and $a_{max} = 0.024$, either randomly or in an ordered manner, to mimic the distinct dynamics of real neurons \cite{Cembrowski2019} within a restricted range ensuring a maximum difference in natural frequencies.
Our results depend on how the natural frequencies of neurons are distributed, determined by their parameters $a$. Additionally, for all simulations, we fix parameters $b=0.2$, $c=-50.0$, and $d=2.0$ \cite{Izhikevich_model_2003, izhikevich2007_book} to maintain the timescale and magnitude of variable $v_i(t)$ around the expected values observed in brain neurons. Random initial conditions for the neurons are used.

For the numerical integration of the equations, we use a fourth-order Runge-Kutta method in C, Python, and Julia programming languages, with consistent results for all algorithms. The time step used was adaptive with a maximum value of $0.1$. Also, for better numerical accuracy near the reset, we interpolate all the variables in a \textit{callback} routine when any neuron variable $v_i(t)$ reaches the threshold $30 \ mV$. 

\section{\label{sec:isolated_dynamics} Individual Neuronal Dynamics}

\begin{figure}[htb!]
    \centering
    \includegraphics[width=0.5\columnwidth]{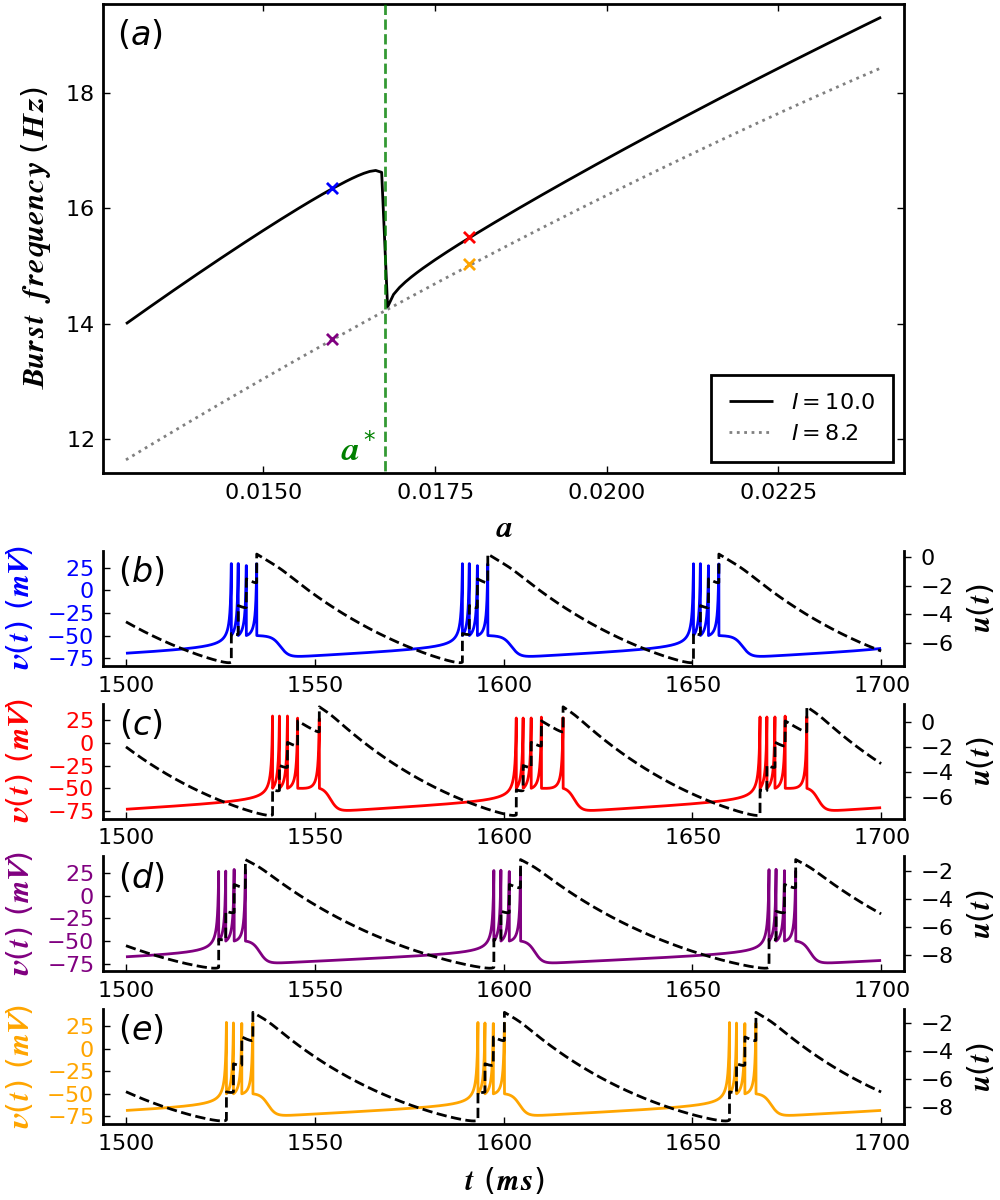}
    \caption{Burst frequency (inverse of the bursting period) and time evolution of the membrane potential $v(t)$ for different values of the bifurcation control parameter $a$ and synaptic input $I$ of the Izhikevich neuron model ($b=0.20$, $c=-50.0$, $d=2.0$).  In panel (a), we show the frequency dependence on $a$ for two relevant values of the synaptic input $I$; in solid black for $I=10.0$ and in dotted gray for $I=8.2$. The values $I=10.0$ and $I=8.2$ represent the constant input for the uncoupled neurons.
    The bottom panels depict the representative time evolution for the isolated neurons near the bifurcation, with parameters: (b) $a=0.016$ and $I=10.0$; (c) $a=0.018$ and $I=10.0$; (d) $a=0.016$ and $I=8.2$; and (e) $a=0.018$ and $I=8.2$. 
    The bursting dynamics displays one additional spike per burst for $a > a^* = 1.678 \times 10^{-2}$ when $I=10.0$, which is followed by a sudden decrease in frequency \cite{marghoti2022intermittent}. 
    }
    \label{fig:time_evolution_frequency_izhikevich}
\end{figure}

First, we present an analysis of isolated neuron dynamics ($\gamma = 0.0$). In Fig.~\ref{fig:time_evolution_frequency_izhikevich}, we show the time evolution and burst frequency response of the isolated Izhikevich neuron as a function of the control parameters $a$ and $I_b$.  The smaller panels show the typical time evolution of neurons near the period-adding bifurcation, for $I_{b}=10.0$ and $I_{b}=8.2$.

For $I_{b} = 10.0$ the neuron model presents a bifurcation that delimits two different periodic dynamics.
A spike gain marks the bifurcation as $a$ increases, which imposes a sudden decrease in burst frequency at the bifurcation value $a^* = 1.678 \times 10^{-2}$.

The cause of such a frequency decrease around $a=a^*$ is related to the fact that each burst needs to be longer to support one additional spike, and the variable $u(t)$ is set to a higher value after the fifth spike. This implies that the time to recover a value likely to generate the next burst is larger, as shown in panels (b) and (c) of Fig.~\ref{fig:time_evolution_frequency_izhikevich}.

The frequency response for $I_{b} =8.2$ is relevant because this is the effective mean current value $I(t)$ that neurons experience in the network when $I_b=10.0$ and $\gamma=0.03$.
Panels (d) and (e) of Fig.~\ref{fig:time_evolution_frequency_izhikevich} show the time evolution of $v(t)$ and $u(t)$ for the same values of parameter $a$ as in panels (b) and (c), respectively.
If $I_{b} =8.2$, the additional fifth spike that characterizes the bifurcation is not present due to the lower rate of change on variable $v_i$, so the input bias $I_b$ can not outweigh the inhibiting effect of term $-u_i$ in Eq. \ref{eq:v_izhik} a fifth time, what prevents the bifurcation for such input bias value and parameter $a$ range.
In a way, for this smaller stimulus bias value, the neuron tends to retain the dynamics of four spikes per burst, despite the similarity of burst frequencies for the same value of $a$.

The bifurcation process alters the sensitivity of neurons to external coupling inputs so that neurons with $a > a^*$ exhibit less frequency variation than those with $a < a^*$ as $I(t)$ changes. This mechanism can promote or inhibit frequency synchronization in our neuronal network, leading to two-phase synchronization regimes.

\section{\label{sec:net_dyn}Collective Neuronal Dynamics}

In a recent study~\cite{marghoti2022intermittent}, the authors explored a similar configuration of heterogeneous neuronal networks exhibiting bistable or intermittent synchronization patterns. Here instead, we investigate how individual neuron dynamics affect the network intermittent dynamics transition times.

\subsection{Network Model Dynamics}

\begin{figure}[htb!]
    \centering
    \includegraphics[width=0.5\columnwidth]{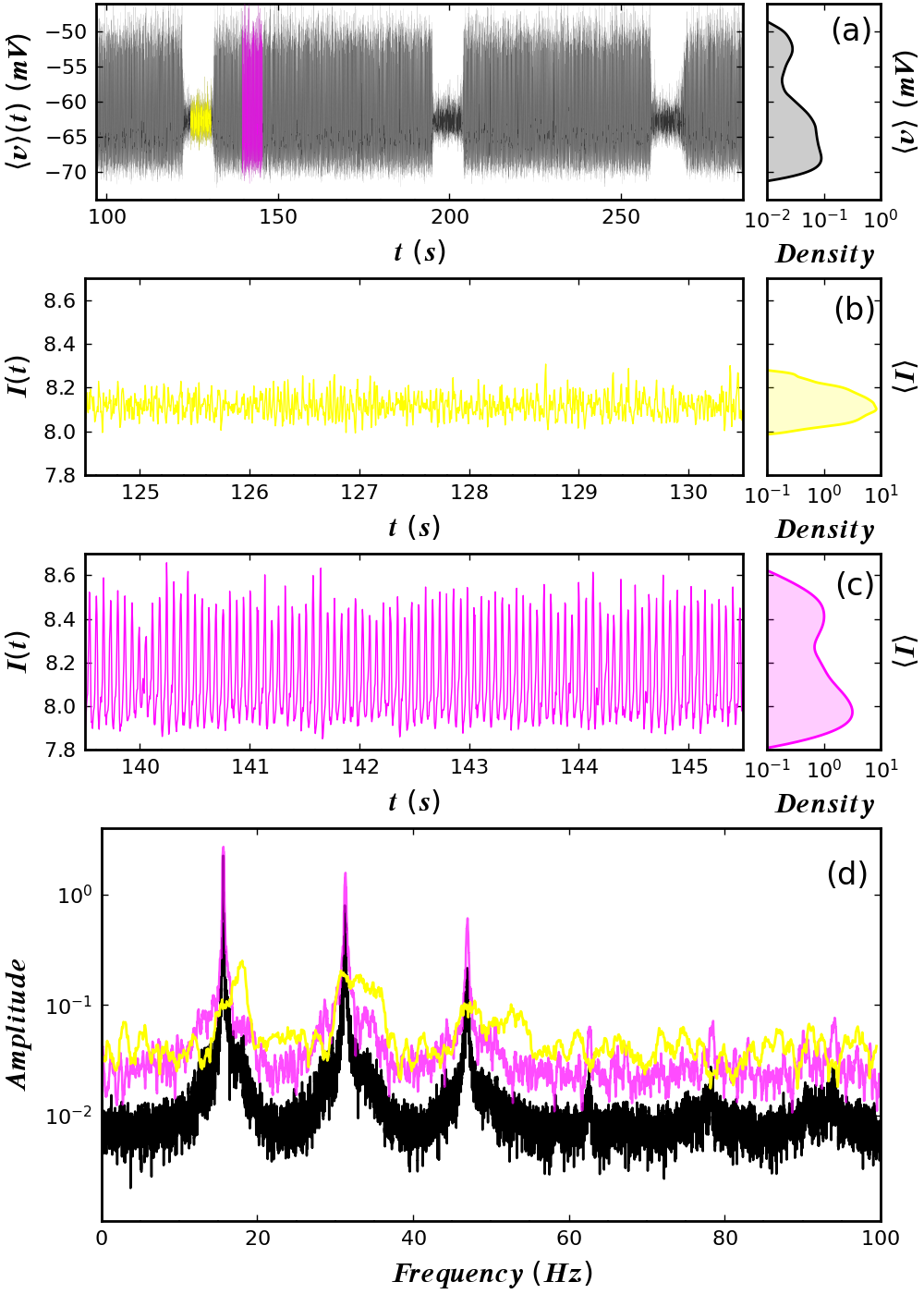}
    \caption{Panel (a) shows the time evolution of the mean-field during intermittent dynamics and the mean-field distribution. Panels (b) and (c) display representative synaptic inputs (as described in Eq.~\ref{eq:I}), collected at periods indicated in panel (a), representing unsynchronized and partially synchronized states, respectively. Panel (d) shows the FFT of the signals using a sampling rate of 200 $Hz$ and a duration period of 200 $s$. The intermittent section is represented in black, the unsynchronized state in yellow, and the partially synchronized state in magenta.}
    \label{fig:mean_field_density}
\end{figure}

We consider the network dynamics, where Fig.~\ref{fig:mean_field_density} displays the time evolution of the network mean-field for intermittent transition dynamics. Our analysis focuses on a network with $N=60$ neurons, a coupling of $\gamma=0.03$, and an input bias value of $I_b = 10.0$. These parameters provide more frequent transition times. Panel (a) depicts some synchronization transitions. The lower amplitude oscillations correspond to the unsynchronized state, while the higher amplitude oscillations represent the partially synchronized state. 
The synaptic inputs for the unsynchronized and partially synchronized states are shown in panels (b) and (c).

On the right side of Fig.~\ref{fig:mean_field_density}, the probability density of each signal is plotted. Notably, the partially synchronized state exhibits an amplitude approximately ten times greater than the unsynchronized mean-field. We present a fast Fourier transform (FFT) of the three signals in panel (d).

Let us highlight the features of the synaptic input signals. First, these signals fluctuate around $8.2$ rather than the value set for $I_b$, which is $10.0$. Second, the amplitude and frequency spectrum of $I(t)$ for the unsynchronized and partially synchronized states differs. Third, the predominant frequency in all signals is around the expected frequency of bursts of neurons, which is near 12 $\sim$ 18 $Hz$.

The formation of intermittent synchronized clusters results from the interplay of two contrasting factors. One is associated with the distinct individual dynamics of neurons inducing an unsynchronized state. The other refers to the influence of the network mean-field to impose phase-synchronized dynamics. The network contains partially synchronized and unsynchronized states, though not globally stable, leading to intermittent transitions between them.

To assess whether the system is phase synchronized or unsynchronized, we use the Kuramoto order parameter \cite{kuramoto2003chemical}. This quantifier for the cluster $g$ with $n_g$ oscillators is given by

\begin{equation}
    R_g(t) = \left| \frac{1}{n_g}\sum_{j=1}^{n_g}e^{i\theta_{j}(t)} \right|,
\end{equation}
where the angle $\theta_{j}(t)$ for each neuron $j$ is determined by
\begin{equation}
    \theta_{j}(t)=2\pi k_{j}+2\pi \frac{t-t_{k-1, j}}{t_{k, j}-t_{k-1, j}},
\end{equation}
and $t_{k,j}$ is the time when the $k$th burst of the $j$th neuron occurs.  If $R_g(t)$ is close to zero, there is a high mismatch among the phases, and for values close to unity, the phases have almost no distinction, and the system is synchronized.
The Kuramoto order parameter for clusters indicates partial phase synchronization when some cluster presents a high value while another presents a low value.

\subsection{Clustered Networks Dynamics}

In this study, we can have a network with partial phase synchronization, where half of its elements form a cluster with the same dynamics. To analyze this, we calculate two order parameters: one for the cluster that can synchronize and another for the remaining neurons. For a network with $N$ neurons, we define $R_1(t)$ as the order parameter for the neuron cluster that never synchronizes and $R_2(t)$ as the order parameter for the cluster that intermittently synchronizes.

\begin{figure*}[htb!]
    \centering
    \includegraphics[width=\textwidth]{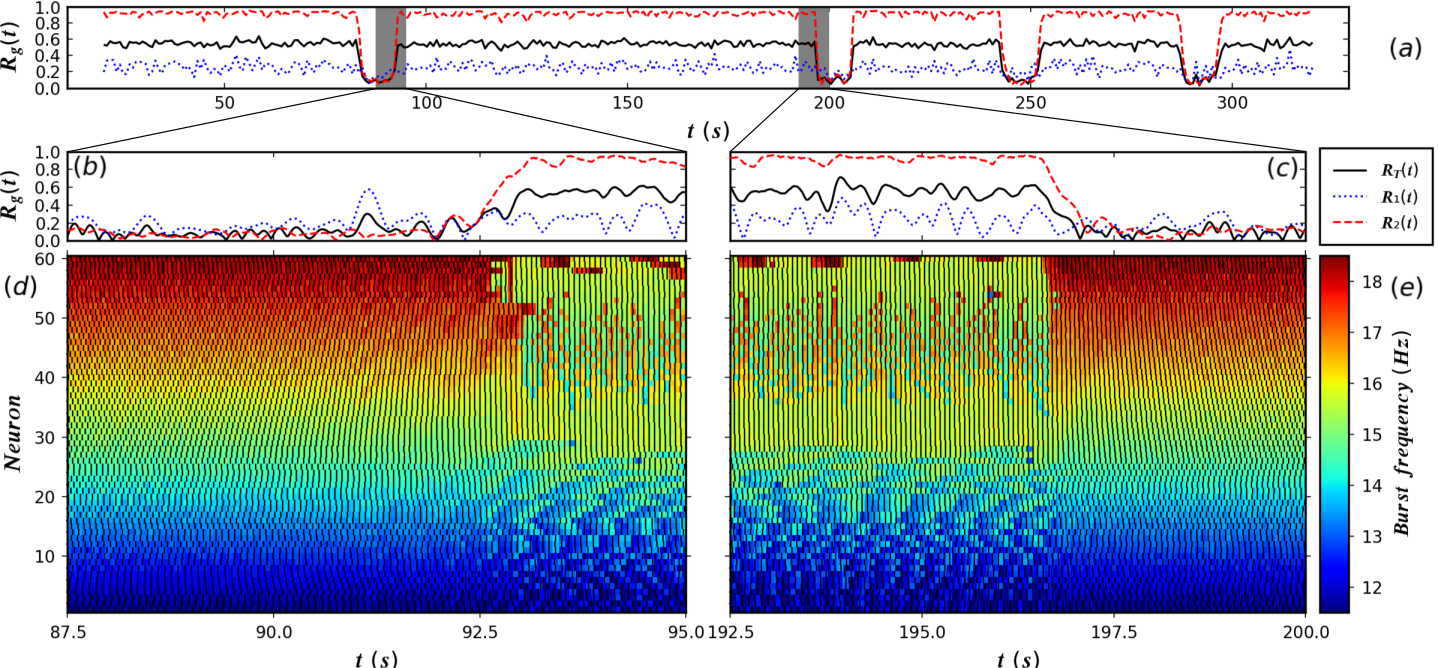}
    \caption{Intermittent transition between unsynchronized and partially synchronized states. We integrate $N=60$ coupled neuronal equations with the following parameters: $b=0.20$, $c=-50.0$, $d=2.0$, $I_b =10.0$ and $\gamma = 0.03$. Additionally, the parameter $a$ is uniformly distributed from $a_{min}=0.013$ to $a_{max}=0.024$ for each neuron 
    We show the Kuramoto order parameter time evolution for (a) several transitions in a larger time scale, (b) the amplification of one transition to a partially synchronized state, (c) one transition to an unsynchronized state, in (d) and (e) the respective raster plots of these transitions.
 	The curves in panel (a) correspond to partial and total order parameters, $R_T(t)$ in solid black for the entire network, $R_1(t)$ in dotted blue for the cluster of neurons with $a<a^*$, and $R_2(t)$ in dashed red for the cluster of neurons with $a>a^*$.
 	In panels (d) and (e) the black slashes are the beginning of each burst and the color scale gives the instantaneous frequency of each burst.
    We notice that the unsynchronized state retains the distinct natural frequencies while the partially synchronized state experiences fluctuations in instantaneous frequencies, eventually leading the network to experience phase synchronization.}
    \label{fig:intermittent}
\end{figure*}

In Fig.~\ref{fig:intermittent}(a), we display the Kuramoto order parameter $R_T(t)$ for the entire network. Furthermore, we delineate the order parameters $R_1(t)$ and $R_2(t)$ for two clusters: the former for neurons with parameter $a < a^*$, and the latter for neurons where $a > a^*$. Panels (b) and (c) provide a closer view of two transitions. One corresponds to the partially phase synchronized state and the other one to the unsynchronized state. Finally, panels (d) and (e) present a raster plot of bursts with the instantaneous burst frequencies represented in the color scale.

The partial phase synchronized state is characterized by a $R_1(t)$ close to 0 while $R_2(t)$ approaches 1. That is, cluster 2 is synchronized, while cluster 1 is not. This state arises due to the preservation of distinct natural bursting frequencies in the neurons, even when subjected to the unsynchronized synaptic input, which has a relatively low amplitude. 
In contrast, the synaptic input in partially synchronized states has a higher amplitude due to the coherence of bursts. This allows neurons in cluster 2 to access different burst frequencies through their time evolution, thereby facilitating their synchronization.

The synaptic input $I(t)$, which couples neurons, stabilizes the network in each synchronization state. However, not all neurons can retain coherent cluster dynamics, which is associated with the parameter $a$ for the neuronal dynamics.
Neurons with parameters $a > a^*$ perform with four and five spikes per burst during the synchronized network dynamics.

\begin{figure}[htb!]
    \centering
    \includegraphics[width=0.5\columnwidth]{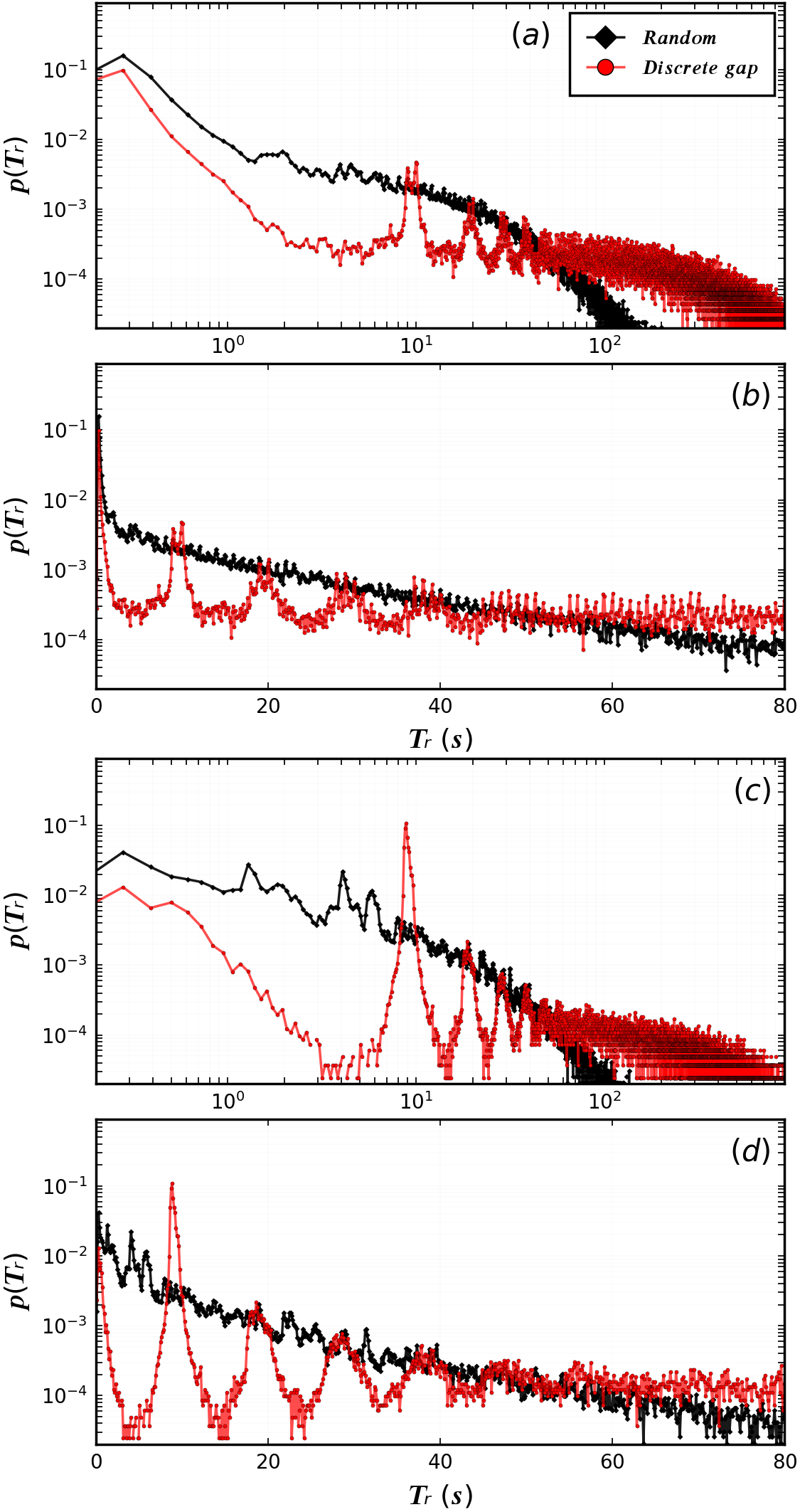}
    \caption{Probability distribution density of residence times for (a-b) partially synchronized and (c-d) unsynchronized states of the network, using random (black) or arranged by discrete gaps (red) of parameter $a$ for $a_{min}<a<a_{max}$.
    The oscillations in $p(T_r)$ arising from frequency gaps due to beat times are discernible when compared to the random scenario.}
    \label{fig:residence_gap_random}
\end{figure}

Fig.~\ref{fig:residence_gap_random} illustrates the probability distributions of residence times in each synchronization state for the random and discrete gap parameters arrangement. Specifically, panels (a) and (b) show the probabilities of residence times for the partially synchronized state, while panels (c) and (d) depict those for the unsynchronized state. Panels (b) and (d) amplify the shorter duration transitions, with the horizontal axis displayed on a linear scale.

There are two different arrangements of parameters, each leading to different residence times. In the case of the uniform random arrangement, we observe an exponential decay in the probability distribution. However, in the orderly scenario, in addition to the exponential decay, we notice peaks with distinguished most probable transition times.
We identify the peaks in the probability distribution of residence times as characteristic transition times between synchronization states. These characteristic times are consistent for both the unsynchronized and partially synchronized states.

\section{\label{sec:beat_formulation}Beat Frequency Statistics of Many Oscillators}

Now, we introduce the general formula for the possible beat frequencies and the number of oscillators that undertake each beat. Furthermore, we explore the impact of the distribution of neuron frequencies with respect to parameter $a$ on the characteristic transition times. This exploration employs a formalism rooted in the frequency beats of periodic oscillators, capturing the behavior of neurons evolving with either phase coherence or phase incoherence. Interestingly, synchronized neurons with similar phases still display beat frequencies during intermittent transitions.  

We consider a set of $N$ periodic oscillators that have $N$ different frequencies, and we derive its properties of the possible frequency beats. Specifically, if we consider two different oscillators $k$ and $p$ with discretized frequencies, the beat frequency between the two can be expressed as follows
\begin{equation}
    f_b^{(k,p)} = | f _k - f _p | = |k-p| \Delta f, \quad \quad k,p = 1,\ 2,\ 3,\ ...,\ N,
\end{equation}
where $\Delta f$ is the minimum frequency gap between any pair of oscillators.
The simplest case where this feature could arise is by an intrinsic discretization of the heterogeneity-providing parameter.

The frequency gap is $ \Delta f = (\partial f / \partial a) \Delta a $ and $\partial f / \partial a$ is the rate of change of frequency concerning the heterogeneity-providing parameter $a$. Moreover, $\Delta a$ is represented as $\Delta a = (a_{\text{max}} - a_{\text{min}})/N$, where $a_{\text{min}}$ and $a_{\text{max}}$ signify the minimum and maximum parameters allocated for the uniform distribution, respectively.

The beat period between two distinct neurons $k$, $p$ is given by
\begin{equation}
    T_b^{(k,p)} = \frac{1}{f_b^{k,p}} =\frac{1}{|k-p|} \frac{1}{ \big(a_{max} - a_ {min}\big) \frac{\partial f}{\partial a}} N,
    \label{eq:T_b_j}
\end{equation}
 where $T_b^{(k,p)}$ assumes a maximum value when the neurons minimally differ by one gap $\Delta f$. In this case, the maximum beat time is given by
\begin{equation}
    T_b = \frac{1}{ \big(a_{max} - a_{min}\big) \frac{\partial f}{\partial a}} N = \frac{1}{\Delta f}.
    \label{eq:T_b}
\end{equation}

Moreover, $T_b$ refers to the time beat between neighboring pairs of oscillators, separated by a gap $\Delta f$. In the case of neurons with arbitrary separations $j = |k-p|$, we have the relation $ T_b^{(j)} = T_b / j$. Even though there are several beat times, we show in Appendix A that time $T_b$ is preferential over any other beat time.

\section{\label{sec:transitions}Transitions Induced by Beat Fre\-quen\-cy}

\begin{figure*}[htb!]
    \centering
    \includegraphics[width=\linewidth]{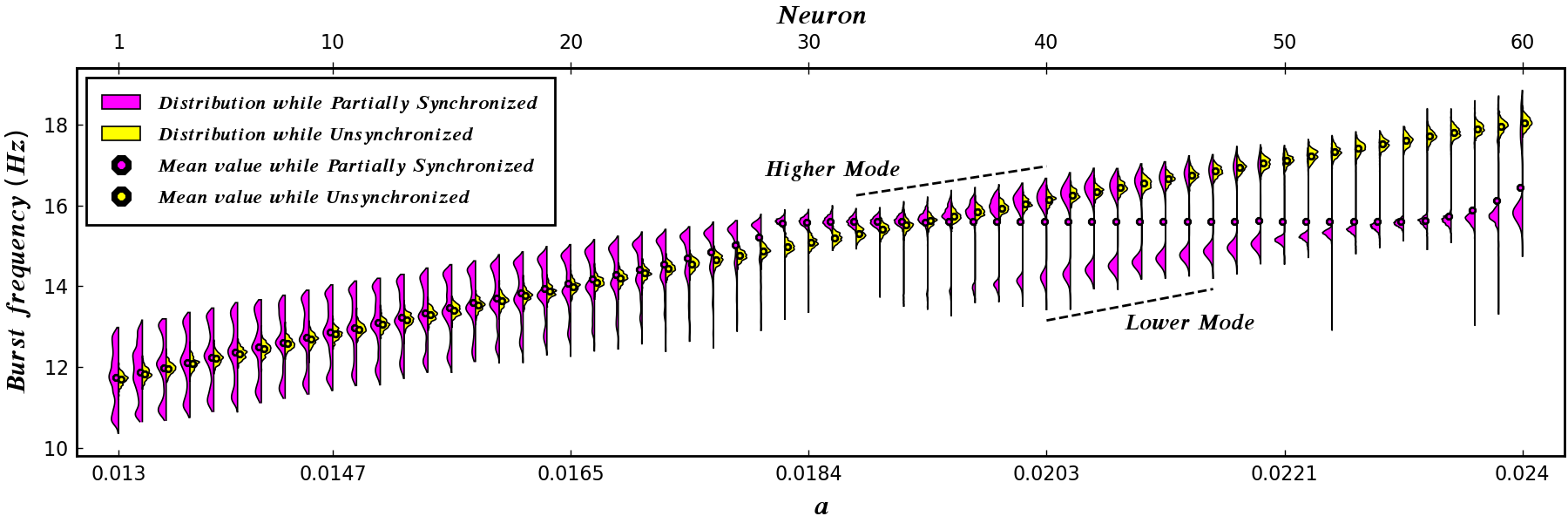}
    \caption{Burst frequency distributions of each neuron in the network. We present the partially phase-synchronized state in magenta and the unsynchronized state in yellow. The network comprises 60 distinct neurons with uniform parameter allocation between $a_{min}=0.013$ and $a_{max}=0.024$, coupling strength $\gamma=0.03$ and current bias $I_b=10.0$. Color shades show the burst frequency distribution, while dots indicate the mean frequency of each neuron. In the unsynchronized state, the mean frequency increases with $a$, while in the partially synchronized state, the synchronized neuron cluster has the same mean but different most probable frequencies due to the bimodal distribution. } 
    \label{fig:freq_neurons_violin}
\end{figure*}

Here, we discuss the source and algebraic agreement with the simulation of beat induced transitions. In Fig. ~\ref{fig:freq_neurons_violin}, we analyze the burst frequency distributions $f$ of each neuron in the network, both in the partially synchronized (magenta) or unsynchronized (yellow) state. For the unsynchronized state, we observe a monotonic increase in the mean frequency as $a$ increases. However, for the partially phase synchronized state, the synchronized neurons share the same mean frequency but there are two modes of more probable frequencies. The slopes of the frequency modes are used to indicate which one of the neuron cluster controls the intermittency induced by the beat frequency.

In Table~\ref{table:Deltaf}, we show the linear fitting for the rate of change $\partial f / \partial a$ of each neuron cluster and for the isolated neuron dynamics ($\gamma = 0.0$).
When the frequency distribution is unimodal, the fit follows the mean frequency, while when the distribution is bimodal, the fitting is done by the mean of two frequency modes (shown in Fig.~\ref{fig:freq_neurons_violin}).
Using these values, we can infer the cause of the transitions regulated by beat frequencies. 

\begin{table}[htb!]
	\footnotesize
\caption{Frequency gaps for the uncoupled network ($I_b=10.0$; $\gamma=0.0$) and the coupled network ($I_b=10.0$; $\gamma=0.03$). In the coupled network, the unsynchronized and partially synchronized frequency gaps for each cluster are presented.}
    \centering
    \label{table:Deltaf}
    \begin{tabular}{c c c c c}
    \toprule
 Network  & Neuron & Pertinent & Fit & Mean $f$\\
 State & Cluster &  Frequencies & $\partial f / \partial a$ & Deviation \\
\cmidrule(lr){1-5}
\multirow{2}{*}{Uncoupled} & 1 & Mean  & $791.62$ & -\\
                           & 2 & Mean  & $627.89$ & -\\

\cmidrule(lr){1-5}
    \multirow{2}{*}{Unsynchronized} & 1 & Mean & $ 662.31 $ & $16.29\%$\\
	                            & 2 & Mean & $ 562.45 $ & $1.32\%$\\
\cmidrule(lr){1-5}
              & 1 & Mean & $726.45$ & $27.45\%$\\
    Partially         & 2 & Mean & $63.65$ & -\\
   Synchronized    & 2 & Higher Mode & $511.32$ & $10.29\%$\\
        & 2 & Lower Mode & $552.60$ & $3.05\%$\\
\bottomrule
    \end{tabular}
\end{table}

\begin{figure}[htb!]
    \centering
    \includegraphics[width=0.5\columnwidth]{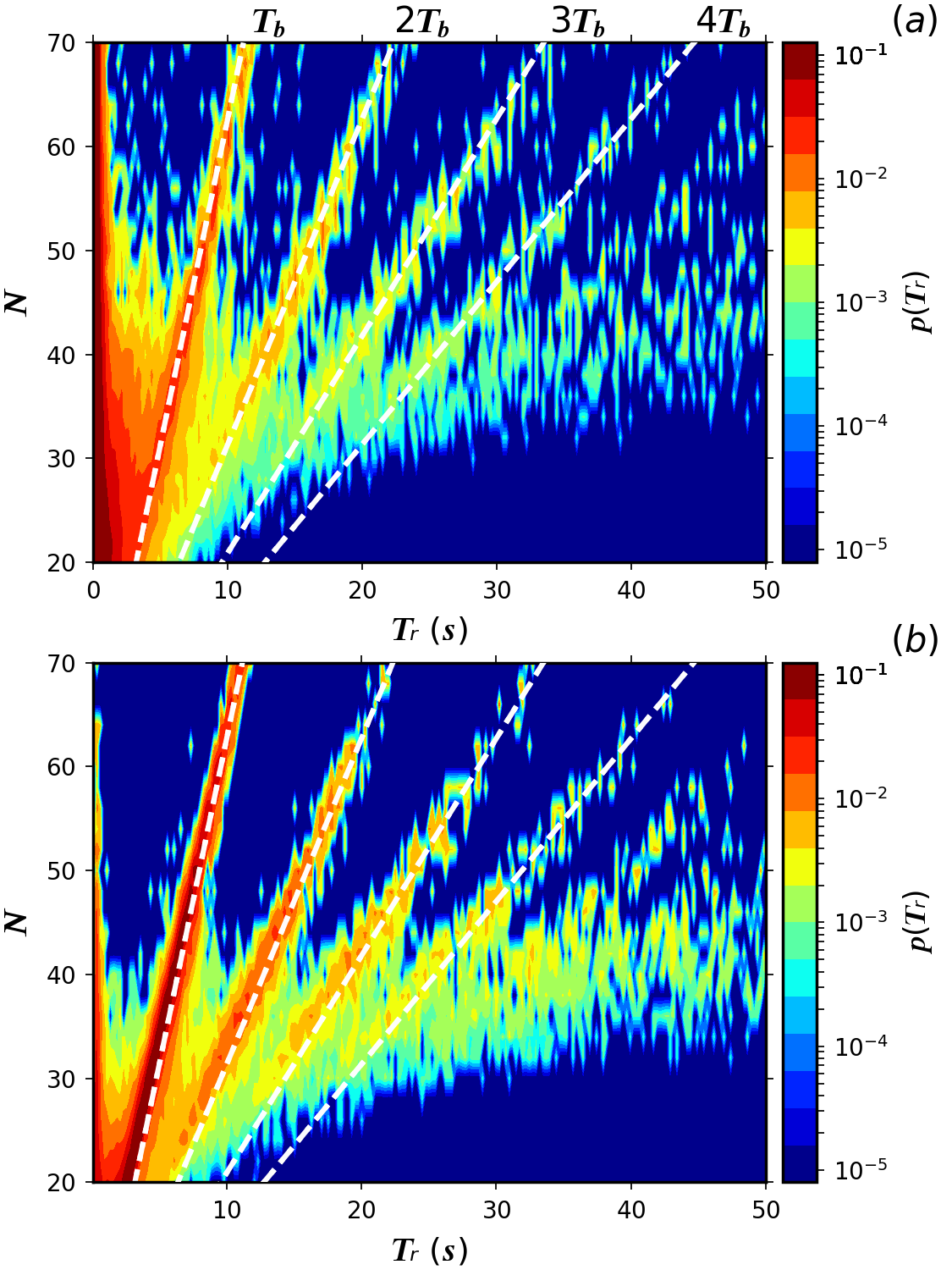}
    \caption{We show, in color scale, the probability density distribution of residence times for (a) partially synchronized and (b) unsynchronized states as a function of the residence times and the number of neurons. The dashed lines are the multiples of the beating period given by Eq.~\ref{eq:T_b} that better fit the peaks of the distributions concerning the frequency response $\partial f / \partial a$. The best fit for both states is for $\partial f / \partial a = 570.0$, this value being similar only to the most probable frequency response to heterogeneity in the cluster that phase-synchronizes.}
    \label{fig:fit_residence_distrib}
\end{figure}

To enable the beat frequency formalism to forecast the peaks in the probability distribution of residence times in both synchronization states, the frequency response to the parameter $a$ provided by heterogeneity should correspond to the observed value of $\partial f / \partial a = 570.0$. In Table~\ref{table:Deltaf}, we show that cluster 2 has the mean frequency deviation of $1.32\%$ for the unsynchronized state. Moreover, for a partially phase synchronized state, the higher mode displays a deviation of $10.29 \%$ and the lower mode a deviation of only $3.05 \%$. Cluster 1 is unlikely to be responsible for the transitions given by beat frequency. The frequency deviations from the correct value are $16.29 \%$ for the unsynchronized state and $27.45 \%$ for the partially phase-synchronized state.

Transition times within cluster 2 are influenced solely by its intrinsic neuron beat frequency, rather than being triggered by other neuron clusters. A key finding of our study is that the frequency gaps are linked to distribution frequency modes, rather than the mean frequency, allowing transition times to be controlled by the beat frequency even for neurons with elevated synchronization.

Fig.~\ref{fig:fit_residence_distrib} shows in color scale the logarithm of the probabilities to observe synchronized or unsynchronized states as a function of the residence times and the number of neurons.
The lines in dashed white correspond to the best fit for beat times $T_b$ (Eq.~\ref{eq:T_b}) and its multiples for the rate $\partial f / \partial a$. 

The above observations agree with the frequency decomposition analysis of the synaptic input (in Fig.~\ref{fig:mean_field_density}), where no frequency with the order of the expected frequency beat has been observed.
This indicates that the network influence through the coupling over neurons does not dictate the transition times related to the peaks in the probability distribution.

In this context, the synaptic input serves as the driving force behind transitions, leading to the emergence of the power law trend in the probability distribution of residence times. Simultaneously, the beat frequency mechanism operates in conjunction with the non-uniformity of the synaptic input to facilitate transitions in intermittent dynamics. However, when the beat frequency mechanism is present, the beat time $T_b$ stands out as a more probable event.

Overall, the limited possibility of burst frequencies for neurons is one of the key requirements for dynamics with preferred transition times. This result emphasizes how the systematical allocation of individual properties impacts the network's complex dynamics through timed self-organization to induce synchronization or desynchronization.

\section{\label{sec:conc}Conclusions}

Intermittent synchronization of neurons occurs when coherent behavior suddenly emerges in sets of incoherent oscillators and is regarded as a key capability in complex neuronal network processing. In this study, we observe this phenomenon in a heterogeneous neuronal network, where the frequency distribution for neurons leads to an unprecedented way to control transitions between synchronization states through beat frequency.

The proximity to the bifurcation parameter value allowed for an additional frequency mode with a spike gain in each burst. This effect enabled neurons that would otherwise be asynchronous to synchronize. As a result, the neurons could maintain properties from their distinct individual dynamics while being phase synchronized, facilitating intermittent transitions at time windows determined by their frequency gaps.

The beating effect increases the probability of transitioning between synchronization states at specific times. We showed how to obtain the preferred transition times in an algebraic form. Furthermore, this phenomenon significantly diminishes in larger networks or non-homogeneous frequency gap distribution. Future work will explore such defects, improving the framework to incorporate different network topologies with more general frequency distributions. Setting out the instances when the beat frequency is a viable mechanism for oscillating networks to control their state transition.

\section*{Acknowledgments}

This work was supported by the Brazilian research agencies Conselho Nacional de Desenvolvimento Cient\'ifico e Tecnol\'ogico (CNPq) Grants Nos. 308441/2021-4, 305189/2022-0, 408254/2022-0 and 140950/2023-0. It has been partially financed by the Coordena\c{c}\~{a}o de Aper\-fei\c{c}oamento de Pessoal de N\'{i}vel Superior, Brasil (CAPES), Finance Code 001, through project No. 88887.833325/2023-00, 88887.898929/2023-00 and 88887.898924/2023-00. Financial support from the Spanish State Research Agency (AEI) and the European Regional Development Fund (ERDF) under Project No. PID2019-105554GB-I00 is also acknowledged.

\section*{Author Contribution Statement}
\textbf{G. M.:} Conceptualization, Investigation, Visualization and Writing - Original Draft. \textbf{T. L. P.:} Supervision, Writing - Review and Editing. \textbf{M. A. F. S.:} Writing - Review and Editing. \textbf{S. R. L.:} Writing - Review and Editing.

\section*{Declaration of Competing Interest}
The authors declare that they have no known competing financial interests or personal relationships that could have appeared to influence the work reported in this paper.

\section*{Appendix A. Beat Times Calculation}
\label{Appendix}

Here, we show which beat time engages the largest number of oscillators. The number of oscillator pairs with a beat time is expressed by $\mathcal{N}\big (T_b^{(j)}\big ) = N-j$, where there are $j$ frequency gaps from $N(N-1)/2$ pairwise oscillators combinations. From Eq.~\ref{eq:T_b_j}, we associate the beat frequency between any two oscillators with the characteristic time $T_b$.
The number of pairs that beat at time $T_b/j$ is given by the sum over the pairs that beat at times $T_b/(k \cdot j)$, which are fractions of $T_b/j$ and presented by

\begin{equation}
	\mathcal{N}(T_b/j) = \sum_{k=1}^{\lfloor (N-1)/j \rfloor} \mathcal{N}(T_b^{(k \cdot j)}) = \sum_{k=1}^{\lfloor (N-1)/j \rfloor} N-jk ,
\end{equation}
for $j = 1$ we have the maximum number of involved oscillators
\begin{equation}
	\mathcal{N}(T_b) = \sum_{k=1}^{N -1} N-k = \frac{N(N-1)}{2}.
\end{equation}

In general, the number of oscillator pairs that beat for $j = 2$ is  $\mathcal{N}(T_b/2) = N (N -2) / 4$, then for $j = 3$ it is $\mathcal{N}(T_b/3)  = N (N-3)/8$, and following this pattern, the number of oscillators that beat in consonance at time $T_B/j$ is given by
\begin{equation}
	\mathcal{N}(T_b/j) = \frac{N(N -j)}{2^j}, \quad j = 1, 2, ... , N-1,
\end{equation}
thus, the fraction of oscillators that beat together at time  $T_b/j$ is written as
\begin{equation}
	\frac{\mathcal{N}(T_b/j)}{N(N-1)/2} = \frac{1}{2^{j-1}}\frac{(N -j)}{(N-1)}.
\end{equation}

Therefore, the discrete and uniform frequency distribution promotes characteristic times $T_b = 1/\Delta f$ that are the most probable events for the network when all the $N(N-1)/2$ possible pairs of oscillators beat together. Subsequent beat times $j+1$ have less than half the number of participating oscillators than the previous beat time $j$; thus, as $j$ increases, fewer oscillators participate in each beat.

\bibliographystyle{model1-num-names}


\end{document}